\documentclass[sigconf,natbib=false]{acmart}
\AtBeginDocument{%
  \providecommand\BibTeX{{%
    \normalfont B\kern-0.5em{\scshape i\kern-0.25em b}\kern-0.8em\TeX}}}

\copyrightyear{2022}
\acmYear{2022}
\setcopyright{rightsretained}
\acmConference[MMSys '22]{13th ACM Multimedia Systems Conference}{June 14--17, 2022}{Athlone, Ireland}
\acmBooktitle{13th ACM Multimedia Systems Conference (MMSys '22), June 14--17, 2022, Athlone, Ireland}
\acmDOI{10.1145/3524273.3532885}
\acmISBN{978-1-4503-9283-9/22/06}

\usepackage{amsmath,graphicx, url}
\usepackage{algorithm}
\usepackage{algpseudocode}
\usepackage{listings}
\usepackage{subcaption}
\usepackage{booktabs}
\usepackage{adjustbox}
\usepackage{xcolor}
\usepackage{url}

\usepackage{breakurl}
\usepackage[style=ACM-Reference-Format,backend=bibtex,sorting=none]{biblatex}
\begin{document}

\title{AQP: An Open Modular Python Platform for Objective Speech and Audio Quality Metrics}

\author{Jack Geraghty}
\email{jack.geraghty@ucdconnect.ie}
\affiliation{%
    \institution{School of Computer Science, University College Dublin}
    \city{Dublin}
    \country{Ireland}
}
\author{Jiazheng Li}
\email{jiazheng.li@ucdconnect.ie}
\affiliation{%
    \institution{School of Computer Science, University College Dublin}
    \institution{Department of Computer Science, University of Warwick}
    \city{Warwick}
    \country{United Kingdom}
}
\author{Alessandro Ragano}
\email{alessandro.ragano@ucdconnect.ie}
\affiliation{%
    \institution{School of Computer Science, University College Dublin}
    \city{Dublin}
    \country{Ireland}
}
\author{Andrew Hines}
\email{andrew.hines@ucd.ie}
\affiliation{%
    \institution{School of Computer Science, University College Dublin}
    \city{Dublin}
    \country{Ireland}
}

\begin{abstract}%
Audio quality assessment has been widely researched in the signal processing area. Full-reference objective metrics (e.g., POLQA, ViSQOL) have been developed to estimate the audio quality relying only on human rating experiments. 
To evaluate the audio quality of novel audio processing techniques, researchers constantly need to compare objective quality metrics. Testing different implementations of the same metric and evaluating new datasets are fundamental and ongoing iterative activities. In this paper, we present AQP -- an \mbox{open-source}, \mbox{node-based}, \mbox{light-weight} Python pipeline for audio quality assessment. AQP allows researchers to test and compare objective quality metrics helping to improve robustness, reproducibility and development speed. We introduce the platform, explain the motivations, and illustrate with examples how, using AQP, objective quality metrics can be (i) compared and benchmarked; (ii) prototyped and adapted in a modular fashion; (iii) visualised and checked for errors. The code has been shared on GitHub to encourage adoption and contributions from the community.
\end{abstract}
\maketitle


\section{Introduction}
\label{sec:intro}

Predictive models that can estimate the quality of speech and audio signals have been widely adopted to develop, evaluate and monitor multimedia applications. Originally developed for the telecoms industry voice communications systems, they are now applied in the development of codecs \cite{Skoglund2020ImprovingOL}, evaluation of hearing aids \cite{Hines:2011} and monitoring and quality assurance for streaming services like Netflix or tele-meetings such as Google Meet \cite{hines2015visqol} as well as for other audio domains such as speech enhancement and music.

Models have been developed to supplement subjective quality assessment, where groups of people are asked to listen to sample audio and rate the quality, producing a mean opinion score (MOS) that is an aggregated average user rating. The models are referred to as objective models and vary in design based on their application requirements. Some models have been standardised and widely adopted under ITU-T recommendations, e.g. PESQ~\cite{rix2001perceptual} and POLQA~\cite{beerends2013perceptual} for speech and PEAQ~\cite{thiede2000peaq} for audio. Others have been released as open-source tools, e.g. ViSQOL~\cite{chinen2020visqol}. 
These models are full-reference i.e., they use both the degraded signal and the reference signal. More recently, deep learning-based models that only use the degraded signal are gaining popularity \cite{serra2021sesqa,mittag21_interspeech,ragano2021more}. 

The earlier models were developed in either C/C++ for speed or MATLAB for research prototyping and were based on monolithic codebases that were difficult to adapt or extend. Models such as NISQA~\cite{mittag21_interspeech}, CDPAM~\cite{manocha2021cdpam}, SESQA~\cite{serra2021sesqa} and WARP-Q~\cite{Wissam2020} have been developed using standard python libraries. Python has matured and become widely adopted in both research and industry deployment with packages and libraries available to implement many standard audio signal processing~\cite{mcfee2015librosa} and machine learning algorithms~\cite{2020SciPy-NMeth,NEURIPS2019_9015} as well as data wrangling~\cite{reback2020pandas} and visualisation~\cite{Hunter:2007}. The trends towards open science and repeatable research have encouraged sharing of code and datasets on platforms such as Github\footnote{\url{https://github.com}} and Zenodo\footnote{\url{https://zenodo.org}}, allowing results in research publication to be easily reproduced and validated.

Full-reference quality metrics consist of several high-level blocks where each block serves a certain scope. Some common blocks are: (1) preprocessing such as normalizing the input level or aligning the reference and the degraded signal, (2) computing an internal representation of the clean and the degraded signal, (3) calculating a similarity score between the two representations, (4) using a regression model to map the similarity score to MOS. Often, quality metrics have been improved by replacing a certain block that was originally used. Replacing blocks is typically desired given that a certain block might work for some datasets but it does not for others. Also, more challenges arise because recent techniques such as generative speech codecs show new degradations that have to be evaluated with quality metrics \cite{jassim2021warp,voran2021full}.
For instance, the first version of ViSQOL uses a Bark-based spectrogram~\cite{hines2015visqol}, which was then replaced by a gammatone-based spectrogram~\cite{sloan2017objective}. Recently another ViSQOL implementation has been proposed~\cite{chinen2020visqol} where the replaced block is in the alignment stage instead of the spectrogram calculation stage. Similarly, other metrics such as PEAQ exhibited improved versions by replacing some blocks~\cite{delgado2020can}. Tools that allow researchers to quickly prototyping, testing, and comparing different implementations of quality metrics are missing. These operations are generally time-consuming and researchers will benefit from having a pipeline to speed up and improve the reproducibility of their work. 

In this paper, we present an Audio Quality Platform (AQP) software implementation addressing the stated problems -- a centralised, comparable and reproducible pipeline for testing and comparison of predictive audio quality metrics. The platform is available on GitHub\footnote{\url{https://github.com/JackGeraghty/AQP}}. Our pipeline follows recent signal processing trends about the urgent need to provide tools for quick deployment and high reproducibility~\cite{eldar2017challenges}.
Since Python is widely used in the research community, we completed this pipeline fully in Python and only uses minimum Python standardized libraries such as librosa\cite{mcfee2015librosa} and Numpy~\cite{harris2020array}. This design allows the most compatibility for environment setup and the most function interpretability during the research. 
Our pipeline allows researchers to modify, remove or add on the functionalities inside each stage of our pipeline to produce a new combination of different evaluation and validation methods. The pipeline can be easily set up via a simple JSON script. Researchers can complete their research via our pipeline and release their experiment script for better community accessibility and reproducibility. Our pipeline also provides visualisation functionality which allows print out the structure of experiments with flowcharts. To the best of our knowledge, our pipeline is the first open source, free to use platform based on standard python libraries that enable use for test harness and model prototyping.

\section{Background}

\subsection{Speech and Audio Quality Models} 
Full reference objective metrics compare a representation of the reference and degraded signal to evaluate the differences and estimate the perceived quality of the degraded signal \cite{hines2015visqol,moller2011speech}. 
In this section, we briefly discuss PESQ and WARP-Q which we use as a case study in this paper. 

PESQ was originally designed to work with narrow-band telephone speech and narrow-band speech codecs~\cite{rix2001perceptual} and was the first widely adopted quality metric. The pre-processing part emulates a telephone handset. Signal disturbances are computed and mapped to MOS. PESQ employs an asymmetry weighting that gives more importance to the added disturbances instead of the attenuated parts on the internal representation of the degraded signal. 

WARP-Q is a the state-of-the-art full-reference metric that uses dynamic time warping cost for MFCC speech representations \cite{jassim2021warp}. WARP-Q was designed for waveform matching, parametric and generative neural vocoder based codecs as well as channel and environmental noise,. WARP-Q shows better performance in correlation, codec quality ranking and versatile general quality assessment than traditional metrics \cite{jassim2021warp}. 

\subsection{Software Tools for Reproducible Open Research}
Structured machine learning platforms are playing important role for open researches nowadays, it allows research community develop their experiment in a standard environment. Basic machine learning libraries like Scikit-learn~\cite{scikit-learn} has motivated fundamental machine learning researchers to carry out easily with shared code bases. More recently, the development of advanced model-based libraries such as Hugging-face Transformer~\cite{wolf-etal-2020-transformers} and OpenCV~\cite{opencv_library} have benefited researchers in natural language processing and computer vision. We can see the value of such standard platforms for audio quality research.

Python and MATLAB have been widely used as a source programming language for signal processing research and application development, which have been proved with state-of-the-art models like WARP-Q~\cite{jassim2021warp}, ViSQOL~\cite{hines2015visqol}. Signal processing research can be carried out at a higher level thanks to the improvement of basic scientific libraries such as librosa~\cite{mcfee2015librosa}, Numpy~\cite{harris2020array} and Matplotlib~\cite{Hunter:2007}. We see a Python based speech and audio processing pipeline in the area of speech and audio signal processing as an important contribution to open research.

A software design pattern is so-called a general repeatable solution to a commonly occurring problem in software design~\cite{10.5555/1076324,10.5555/186897}. The modular components in our designed pipeline enable easy configuration and visualisation. The design of each functional modular also supports developers to test various models in parallel. These advantages of our software structure are so-called creational design patterns and structural design patterns~\cite{10.5555/186897}. Standardized input and output API setup allows developers to reuse components in the pipeline for duplicate tasks~\cite{10.5555/186897}. For example, the same processing methods can be applied to both reference and degraded signals. It also enables developers recreation when situations like changing filter banks.
 
Benefited from design patterns, this pipeline structure improves the interpretability -- researchers can easily compare the results from different components to debug or try with various setups. It also increases the maintainability -- allows developers to keep different versions of experiments and easily validate/test them without structural redesign. Such design enables comparability -- users can simply set up to leverage parallelisable parts. Such structure allows the research community hands-on easily and friendly for basic-level understanding and reproduction of experiments, it also assists with higher-level customisation and results re-creation.
 
The expectation of reproducible research has gained traction within the signal processing community~\cite{4218335}. Some ACM conferences, including MMSys, offer different badges that indicate the level of repeatability, reproducibility and replicability of a submission\cite{badging}. The badges are awarded upon a review of the submission. We anticipate a similar boom in the use of standardised pipeline structure for reproducibility as recently seen in other domains that have adopted machine learning\cite{wolf2020huggingfaces,opencv_library}.  

\section{AQP Platform}
\label{sec:platform}

The following section describes the architecture of the platform and how it is utilized to provide the high level of modularity required. Then a description of the core nodes of the pipeline and their use cases are given. These are: the \texttt{LoopNode}, for looping over data, the \texttt{EncapsulationNode}, for managing sub-pipelines and the \texttt{SinkNode}, for managing the control flow of the pipeline. The section concludes with an overview of how the pipeline is executed, configured and visualized.

The core architecture of the pipeline is built upon the Directed Acyclic Graph (DAG) data structure. A DAG is an implementation of a directed graph, but with no directed cycles. The acyclic property of the graph is used to prevent infinite loops occurring during the execution of the pipeline, without it, explicit control flow logic would need to be required to prevent these loops from occurring. The node based structure of the DAG is used to provide the high level of modularity required by the pipeline. Each node in the graph is used to encapsulate some logic that will be performed on signal data. Functionality is encapsulated as nodes into logical blocks. By re-ordering/introducing/removing nodes from the pipeline, different configurations can be tested without having to modify the codebase.

Data is passed through the pipeline using a Python dictionary. When a node is being executed it can retrieve data from the dictionary, add/update data already in the dictionary or do nothing to it. This method increases the modularity of the pipeline, as no node requires knowledge of the rest of the pipeline. This also improves the testability of nodes, a node can be tested in isolation as the only requirement is a dictionary, which can be mocked easily.

\textbf{Base Node:} 
All nodes in the pipeline inherit from the \textbf{\textit{Node}} base class. This class contains the common properties for all nodes and indicates that all nodes must implement the \textbf{\textit{execute}} function. All nodes have at least two required fields, an ID (for connecting a node with its children) and a type (for debug purposes), as well as two optional parameters, \textit{output\_key} and \textit{draw\_options}. The \textit{output\_key} is used for assigning the result calculated back to the result dictionary. The \textit{draw\_options} are used to provide additional arguments for drawing a node when creating a DOT file representation for the graph. 

\textbf{Loop Node:}
Audio quality metrics often deal with multiple audio channels being active for a given signal. All of these channels need to be evaluated in some form before a final value can be associated with the signal. The functionality used is identical for each channel, so the ability to loop over channels (or some other variable) is required. The \texttt{LoopNode} is used to achieve this functionality. When creating a \texttt{LoopNode} a sub-pipeline of nodes is defined and then during the execution of that node an iterable (e.g. list of active channels) is retrieved from the result dictionary. The sub-pipeline is executed for each entry in the iterable, with the result of each iteration being stored in a separate dictionary. The final dictionary is then assigned to the main result.

\textbf{Encapsulation Node:}
This node allows nested configurations where a pipeline in a JSON file that can be used within another graph as part of a larger pipeline. This allows for a large group of nodes, e.g. the main ViSQOL functionality, to be grouped into a single file and reused in other configurations without having to redefine it. When the \texttt{execute} function of the node is called, it executes all of the nodes contained within the pipeline of the \texttt{EncapsulationNode}. Apart from the reusability the use of the \texttt{EncapsulationNode} also serves as a method of keeping pipeline configuration files succinct and well organised, simplifying for abstraction and visualisation. 

\textbf{Sink Node}: Comparing different configurations in a single pipeline is represented as a node having multiple children, where each branch indicates a different configuration to evaluate. After each of these branches have been evaluated the results of each branch need to be available in a single, shared node for further use. The \texttt{SinkNode} offers the functionality to collect a set of results from different branches and prevent further execution of the graph until all these results are collected. Until the \texttt{SinkNode} has seen the expected number of results to collect it will return \textit{None}, indicating to the pipeline to evaluate a different branch before proceeding with the children of the \texttt{SinkNode}. Once it has seen the expected number of results, execution of the pipeline continues as normal.

\textbf{Pipeline Execution:}
Nodes in the pipeline get executed in a depth first manner. Each node maintains a list of all its children. Provided the execution of a node was successful, then all of its children are executed next. The \texttt{execute} function of a node should return the result dictionary if execution was successful and \texttt{None} otherwise. If the return value is \texttt{None}, then the current node's children do not get evaluated, instead the next node on the current level of the DAG is executed. Pseudocode for the execution of the pipeline is given in Algorithm \ref{alg:traversal}.

\begin{algorithm}
\caption{Modified Depth First Traversal}\label{alg:traversal}
\begin{algorithmic}
\Require $node, result $
\State $stack \gets []$
\State $stack.append(node)$
\While{$len(stack) > 0$}
    \State $current\_node \gets stack.pop()$
    \State $children \gets current\_node.children$
    \If{$(r \gets current\_node.execute(result))\: is\: None$}
        \State $continue$
    \EndIf
    \For{$i \gets len(children)-1, 0$}
        \State $stack.append(children[i])$
    \EndFor
\EndWhile
\end{algorithmic}
\end{algorithm}

\textbf{Configuration:}
The pipeline is defined through JSON. Each node in the DAG corresponds to an entry in the JSON file. This entry contains all the necessary parameters for creating a node (children, output\_key, etc.). An entry's key is used as the ID for the node it is describing. On startup the configuration file is loaded, deserialized to node objects, and then the DAG is constructed using the root node and then children of each node. Defining the pipeline through JSON enables the testing of different pipeline configurations without having to touch the codebase of the pipeline.

\textbf{Visualization:}
Having a visual representation of the pipeline is useful for both debugging and conveying what steps are involved in the pipeline. AQP creates a DOT file for the pipeline created by using NetworkX \cite{networkx}. This DOT file can then be used to generate a diagram of the pipeline. The \texttt{draw\_options} parameter of a node can be used to specify how the node should be drawn when the diagram is generated. The pipeline produces a \textit{.dot} file representation of the graph, which can be used to generate a visualization for the graph, an example of which can be seen in Figure \ref{fig:pipeline}. This \textit{.dot} file can also be further modified to produce the desired visualization. 

\section{AQP Example Case Study}
\label{sec:casestudy}
\begin{figure}[th!]
  \centering
  \centerline{\includegraphics[width=9cm]{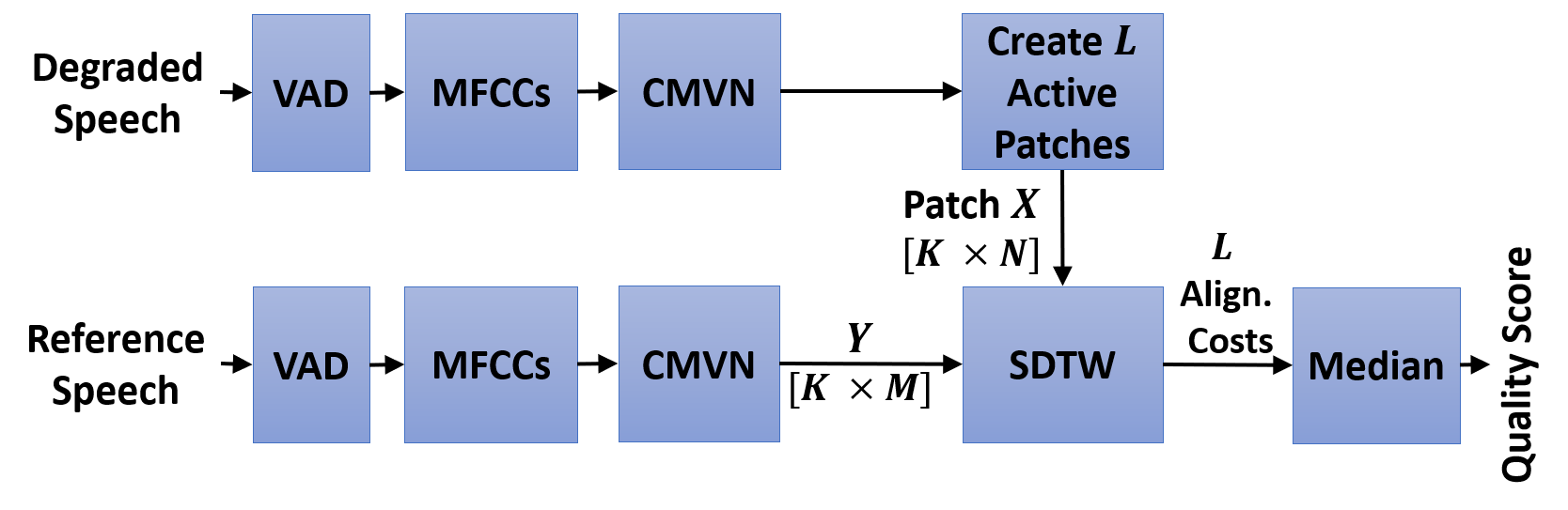}}
\caption{Block diagram of the WARP-Q metric reproduced from \cite{jassim2021warp}. %
}
\label{fig:warpq}
\end{figure}

To demonstrate the prototyping, benchmarking and presentation capabilities of the proposed platform, a case study is presented. Two audio quality metrics, WARP-Q and PESQ, were tested and compared for the GenSpeech dataset. Both metrics were ported and encapsulated in AQP nodes for evaluation within the pipeline's architecture. A variation of the WARP-Q metric was also created, where the node creating the signal features Mel-frequency cepstral coefficients (MFCCs) were substituted with a Mel spectrogram node. This setup illustrates how AQP can be used as an end-to-end research platform for testing and comparing multiple audio quality metrics and for prototyping and comparing performance for different configurations. The evaluation dataset is loaded at the beginning of the pipeline from a CSV file, the quality metric algorithms are then run on the dataset entries with the dataset input being updated with the quality scores and finally, the results are graphed and a LaTeX table generated with the results.
\begin{figure*}[h!]
  \centering
  \centerline{\includegraphics[width=\textwidth]{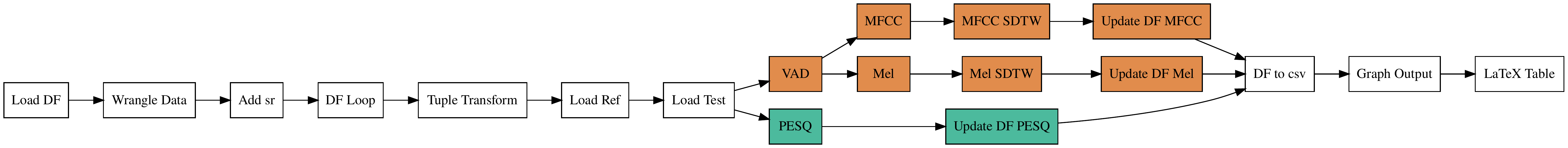}}
\caption{Auto-generated pipeline visualization for the configuration described in Section~\ref{sec:casestudy}. %
}
\label{fig:pipeline}
\end{figure*}

The block diagram seen in Figure \ref{fig:warpq} was used as a reference for re-implementing the WARP-Q metric in the pipeline's architecture. Three AQP nodes were implemented specifically for WARP-Q, the \textit{WarpQVadNode} for identifying the active voice patches, the \textit{MFCCNode} for computing the (MFCCs) and the cepstral mean and variance normalisation (CMVN) of the input signals and the \textit{WarpQSDTWNode}, which performs the SDTW algorithm and calculates the final quality score. These three nodes were then defined within an EncapsulationNode.

By encapsulating the WARP-Q algorithm within a collection of nodes it is possible to add/remove/substitute nodes to evaluate the impact that such a change has on the final quality score produced by the algorithm, e.g removing the VAD. Another benefit is that it is possible to test different variations of the parameters of a node against the base implementation of WARP-Q, or any other quality metric. 

Algorithms that do not need to be changed for experimental purposes can quickly and easily be deployed within the pipeline for bench-marking purposes. For example, the PESQ metric was originally implemented in C and we used a python wrapper of PESQ~\cite{pypesq}. It was not separated into nodes for each of the algorithm's functional blocks but simply wrapped inside a single AQP node so that it could be deployed in the pipeline.

The pipeline used to perform this case study can be seen in Figure \ref{fig:pipeline}. The GenSpeech dataset, which consists of paths to audio signals, codec used and the MOS for the test signal, is loaded into the pipeline in the form of a pandas dataframe. Pairs consisting of a reference signal path and a test signal path are then retrieved from this dataframe and loaded as Numpy arrays using librosa. The signals are then passed to both the WARP-Q nodes and PESQ node where the respective algorithms are run. When each algorithm is finished the final result is then added to the corresponding entry in the dataframe. Once all pairs of signals are processed the dataframe is written to disk and then passed to the output nodes. The output nodes used in this configuration create a simple plot of the sample MOS vs the calculated MOS and then a LaTeX table showing the Pearson and Spearman correlation between the sample MOS and calculated MOS.

\begin{figure}[th!]
  \centering
  \centerline{\includegraphics[width=\linewidth]{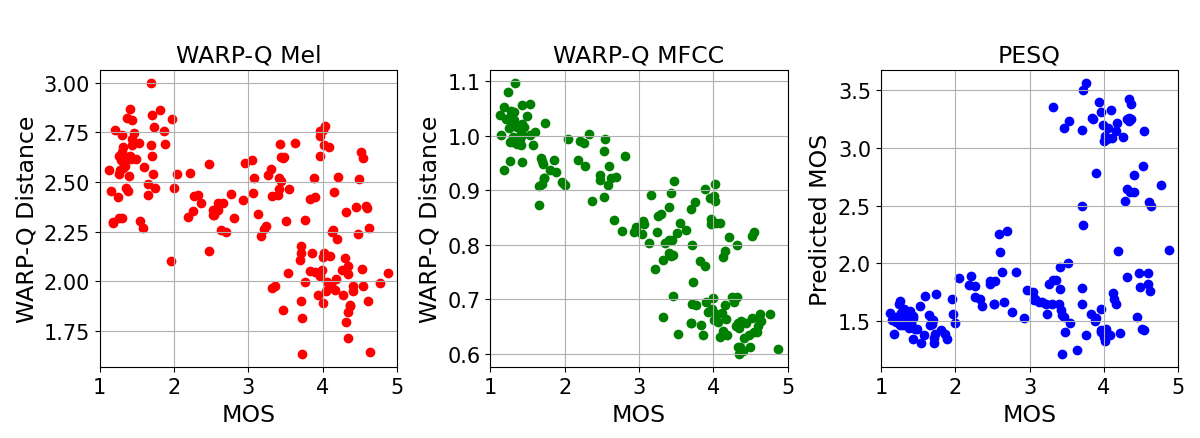}}
  \caption{Sample graph output produced by the pipeline in Fig.~\ref{fig:pipeline}}
  \label{fig:scatter_output}
\end{figure}

The output nodes used to demonstrate that the pipeline can be used as an end-to-end platform, including data visualization and presentation (e.g. Figure~\ref{fig:scatter_output}). Someone using this platform can implement their own output nodes to match their intended use case\footnote{Note that the two quality metrics used in this paper were selected for illustrative purposes. We are not presenting a new or adapted quality metric. The  results comparison between the metrics is only presented to demonstrate the utility of AQP.}.

\section{Discussion and Conclusions}
When developing and testing objective quality metrics many challenges arise. Quality metrics are based on the interaction of many processing blocks. Removing, replacing, and experimenting with new blocks and new datasets are common time-consuming operations. Typically, researchers build their own software platform from scratch which is \mbox{time- consuming} and risks limiting reproducibility, low robustness, and requires time. In this paper, we have proposed a platform able to solve these issues. The proposed platform allows researchers to easily remove/replace blocks and/or datasets. For instance, a cognitive mapping layer can be easily modified (e.g. \cite{Chinen2021dln}), or the impact of new/modified preprocessing steps on that mapping can be evaluated. Thanks to the visualisation tool of the blocks, researchers are able to quickly see how the new block is tied to a previous set of blocks. This ensures that the right model is used with the right feature inputs. Comparing different implementations of the same quality metric is quick and less prone to mistakes. With a simple JSON configuration file, the parameters of a block can be easily modified, without accessing the internal code. Objective quality metrics have exhibited continuous modifications to be adapted to the new audio processing algorithms and applications. By using our platform, someone can quickly prototype the new model with different configurations and testing against other models. For example, recent audio processing techniques that introduce new degradations such as generative speech codecs will benefit from the proposed platform. The presented platform provides a lightweight, standardised method of taking input data and producing outputs for audio quality metric experiments. Future work will involve providing more audio quality metrics and datasets in the base package so researchers can easily test their own quality metrics and different configurations of existing ones (e.g. for spatial audio models \cite{narbutt2018ambiqual} or no reference models \cite{Martinez2019}). Further error checking for pipeline configurations will be introduced. While the examples presented did not rely on machine learning nodes AQP will simplify the integration of machine learning models into objective audio quality models allowing different training scenarios to be deployed and evaluated. 

\label{conclusions}

\section{Acknowledgements}
This publication has emanated from research conducted with
the financial support of Science Foundation Ireland (SFI) under Grant Numbers 18/CRT/6224, 17/RC/2289\_P2 and 17/RC-PhD/3483. For the purpose of Open Access, the authors have applied a CC BY public copyright licence to any Author Accepted Manuscript version arising from this submission.
\printbibliography

\end{document}